\documentclass{article}
\usepackage{spconf}
\usepackage{pgfplots,amsmath,graphicx}
\usepackage{multirow}
\usepackage{cite}
\usepackage{url,afterpage}
\usepackage{enumitem} 

\usepackage{amsfonts}
\usepackage{amssymb}
\usepackage{amsthm}
\usepackage{graphicx, color, transparent}
\usepackage{tabularx}
\usepackage{makecell}
\usepackage{multirow}
\usepackage{pbox}
\usepackage{tikz}
\usetikzlibrary{plotmarks}
\usepackage{pgfplots}
\pgfplotsset{compat=1.7}
\usetikzlibrary{calc}
\usetikzlibrary{shapes,arrows}
\usetikzlibrary{decorations.markings}
\usepackage{amsthm}
\usepackage{cite,flushend}
\usepackage{array}
\makeatletter
\newcommand*\bigcdot{\mathpalette\bigcdot@{.5}}
\newcommand*\bigcdot@[2]{\mathbin{\vcenter{\hbox{\scalebox{#2}{$\m@th#1\bullet$}}}}}
\makeatother

\newcommand{\Enc}{\mathsf{Enc}}
\newcommand{\Dec}{\mathsf{Dec}}

\newcommand*\xor{\mathbin{\oplus}}

\makeatletter
\newtheorem*{rep@theorem}{\rep@title}
\newcommand{\newreptheorem}[2]{%
	\newenvironment{rep#1}[1]{%
		\def\rep@title{\Cref{##1}}%
		\begin{rep@theorem}}%
		{\end{rep@theorem}}}
\newcommand*{\textlabel}[2]{%
	\edef\@currentlabel{#1}
	\phantomsection
	#1\label{#2}
}
\makeatother

\tikzset{XOR/.style={draw,circle,append after command={
        [shorten >=\pgflinewidth, shorten <=\pgflinewidth,]
        (\tikzlastnode.north) edge (\tikzlastnode.south)
        (\tikzlastnode.east) edge (\tikzlastnode.west)
        }
    }
}
\tikzset{line/.style={draw, -latex',shorten <=1bp,shorten >=1bp}}
\title{LOW-COMPLEXITY AND RELIABLE TRANSFORMS FOR\\ PHYSICAL UNCLONABLE FUNCTIONS}
\name{Onur G\"unl\"u and Rafael F. Schaefer\thanks{O. G\"unl\"u and R. F. Schaefer were supported by the German Federal Ministry of Education and Research (BMBF) within the national initiative for ``Post Shannon Communication (NewCom)'' under Grant 16KIS1004.}}
\address{Information Theory and Applications Chair, Technische Universit\"at Berlin\\
		\{{guenlue}, {rafael.schaefer}\}{@tu-berlin.de}\\
		  }%
\begin{document}
\ninept
\maketitle
\begin{abstract}
	Noisy measurements of a physical unclonable function (PUF) are used to store secret keys with reliability, security, privacy, and complexity constraints. A new set of low-complexity and orthogonal transforms with no multiplication is proposed to obtain bit-error probability results significantly better than all methods previously proposed for key binding with PUFs. The uniqueness and security performance of a transform selected from the proposed set is shown to be close to optimal. An error-correction code with a low-complexity decoder and a high code rate is shown to provide a block-error probability significantly smaller than provided by previously proposed codes with the same or smaller code rates.
\end{abstract}

\section{Introduction}
Biometric identifiers such as fingerprints are useful to authenticate a user. Similarly, secret keys are traditionally stored in non-volatile memories (NVMs) to authenticate a physical device that contains the key. NVMs require hardware protection even when the device is turned off since an attacker can try to obtain the key at any time. A safe and cheap alternative to storing keys in NVMs is to use physical identifiers, e.g., fine variations of ring oscillator (RO) outputs, as a randomness source. Since invasive attacks to physical identifiers permanently change the identifier output, there is no need for continuous hardware protection for physical identifiers \cite{pufintheory}.

Physical unclonable functions (PUFs) are physical identifiers with reliable and high-entropy outputs \cite{GassendThesis,PappuThesis}. PUF outputs are unique to each device, so they are used for safe and low-complexity key storage in digital devices. These keys can be used for private authentication, secure computation, and encryption. Replacing such identifiers is expensive, so key-storage methods should limit the information the public data leak about the identifier outputs. Moreover, the same device should be able to reconstruct a secret key generated from the noiseless outputs by using the noisy outputs and public information. The ultimate secret-key vs. privacy-leakage rate tradeoffs are given in \cite{IgnaTrans,LaiTrans,benimdissertation}. The secret-key and privacy-leakage rate limits for a suboptimal chosen-secret (CS) model called \emph{fuzzy commitment scheme} (FCS) \cite{FuzzyCommitment} are given in \cite{IgnatenkoFuzzy}. We consider the FCS to compare different post-processing methods applied to PUFs. Asymptotically optimal CS model constructions are given in \cite{bizimWZ} and similar comparison results can be obtained by using these constructions.

Physical identifier outputs are highly correlated and noisy, which are the two main problems in using PUFs. If errors in the extracted sequences are not corrected, PUF reliability would be low. If correlations are not eliminated, machine learning algorithms can model the PUF outputs \cite{MLPUF}. To solve the two problems, the discrete cosine transform (DCT) is used in \cite{bizimtemperature} to generate a uniformly-distributed bit sequence from PUFs under varying environmental conditions. Similarly, the discrete Walsh-Hadamard transform (DWHT), discrete Haar transform (DHT), and Karhunen-Lo\`{e}ve transform (KLT) are compared in \cite{bizimMDPI} in terms of the maximum secret-key length, decorrelation efficiency, reliability, security, and hardware cost. The DCT, DWHT, and DHT provide good reliability and security results, and a hardware implementation of the DWHT in \cite{bizimMDPI} shows that the DWHT requires a substantially smaller hardware area than other transforms. There are two main reasons why the DWHT can be implemented efficiently. Firstly, the matrix that represents the DWHT has elements $1$ or $-1$, so there is no matrix multiplication. Secondly, an input-selection algorithm that is an extension of the algorithm in \cite{InputSelection} allows to calculate two-dimensional (2D) DWHT recursively. Based on these observations, we propose a new set of transforms that preserve these properties and that significantly improve the reliability of the sequences extracted from PUFs.   

The FCS requires error-correction codes (ECCs) to achieve the realistic block-error probability of $\displaystyle P_\text{B}\!=\!10^{-9}$ for RO PUFs. The ECCs proposed in \cite{bizimMDPI} have better secret-key and privacy-leakage rates than previously proposed codes, but in some cases it is assumed that if multiple bits are extracted from each transform coefficient, each bit is affected by independent errors. This assumption is not valid in general. Thus, we extract only one bit from each transform coefficient. The contributions of this work are as follows.
\begin{itemize}
	\item We propose a new set of 2D orthogonal transforms that have low-complexity hardware implementations and no matrix multiplications. The new set of transforms are shown to provide an average bit-error probability smaller than the most reliable transform considered in the PUF literature, i.e., DCT. 
	\item Bit sequences extracted using a transform selected from the new set of transforms are shown to give good uniqueness and security results that are comparable to state-of-the-art results.
	\item We propose a joint transform-quantizer-code design method for the new set of transforms in combination with the FCS to achieve a block-error probability substantially smaller than the common value of $10^{-9}$ with perfect secrecy. 
\end{itemize}

This paper is organized as follows. In Section~\ref{sec:fuzzycommitment}, we review the FCS. The transform-coding algorithm to extract secure sequences from RO PUFs is explained in Section~\ref{sec:commonsteps}. A new set of orthogonal transforms that require a small hardware area and that result in bit-error probabilities smaller than previously considered transforms is proposed in Section~\ref{sec:neworth}. In Section~\ref{sec:comparisons}, we compare the new transforms with previous methods and show that the proposed ECC provides a block-error probability for the new selected transform (ST) that is smaller than for previously considered transforms.
\section{Review of the Fuzzy Commitment Scheme}\label{sec:fuzzycommitment}
\begin{figure}
\centering
\resizebox{0.93\linewidth}{!}{

\begin{tikzpicture}
\node (a) at (0,-1.5) [XOR,scale=1.0] {};
\node (b) at (6,-1.5) [XOR,scale=1.0] {};
\node (f) at (0,-0.7) [draw,rounded corners = 6pt, minimum width=2.8cm,minimum height=3cm, align=left] {};
\node (g) at (6,-0.7) [draw,rounded corners = 6pt, minimum width=3.25cm,minimum height=3cm, align=left] {};
\node (d) at (0,-0.1) [draw,rounded corners = 6pt, minimum width=2.6cm,minimum height=0.7cm, align=left] {$
C^N = \Enc\left(S\right)$};
\node (c) at (3,-2.7) [draw,rounded corners = 5pt, minimum width=1.3cm,minimum height=0.65cm, align=left] {$P_{Y|X}$};
\node (e) at (6,-0.1) [draw,rounded corners = 6pt, minimum width=2.6cm,minimum height=0.7cm, align=left] {$\hat{S} = \Dec\left(C^N\!\xor\! E^N\right)$};
\draw[decoration={markings,mark=at position 1 with {\arrow[scale=1.5]{latex}}},
    postaction={decorate}, thick, shorten >=0.6pt] (a.east) -- (b.west) node [midway, above] {$W^N$};
\node (a1) [below of = a, node distance = 1.2cm] {$X^N$};
\node (b1) [below of = b, node distance = 1.2cm] {$Y^N=X^N\!\xor\!E^N$};
\draw[decoration={markings,mark=at position 1 with {\arrow[scale=1.5]{latex}}},
    postaction={decorate}, thick, shorten >=1.4pt] (a1.north) -- (a.south);
\draw[decoration={markings,mark=at position 1 with {\arrow[scale=1.5]{latex}}},
    postaction={decorate}, thick, shorten >=1.4pt] (a1.east) -- (c.west);
\draw[decoration={markings,mark=at position 1 with {\arrow[scale=1.5]{latex}}},
    postaction={decorate}, thick, shorten >=1.4pt] (c.east) -- (b1.west);
\draw[decoration={markings,mark=at position 1 with {\arrow[scale=1.5]{latex}}},
    postaction={decorate}, thick, shorten >=1.4pt] (b1.north) -- (b.south);
\node (a2) [above of = d, node distance = 1.5cm] {$S$};
\node (f2) [below of = f, node distance = 2.5cm] {Enrollment};
\node (g2) [below of = g, node distance = 2.5cm] {Reconstruction};
\node (b2) [above of = e, node distance = 1.5cm] {$\hat{S}$};
\draw[decoration={markings,mark=at position 1 with {\arrow[scale=1.5]{latex}}},
    postaction={decorate}, thick, shorten >=1.4pt] (e.north) -- (b2.south);
\draw[decoration={markings,mark=at position 1 with {\arrow[scale=1.5]{latex}}},
    postaction={decorate}, thick, shorten >=1.4pt] (a2.south) -- (d.north); 
\draw[decoration={markings,mark=at position 1 with {\arrow[scale=1.5]{latex}}},
    postaction={decorate}, thick, shorten >=1.4pt] (b.north) -- (e.south) node [midway, right] {$C^N\!\xor\! E^N$};
\draw[decoration={markings,mark=at position 1 with {\arrow[scale=1.5]{latex}}},
    postaction={decorate}, thick, shorten >=1.4pt]  (d.south) -- (a.north) node [midway, left] {$C^N$};;
\end{tikzpicture}
}
 \caption{The fuzzy commitment scheme (FCS).}\label{fig:fuzzycommitment}
\end{figure}
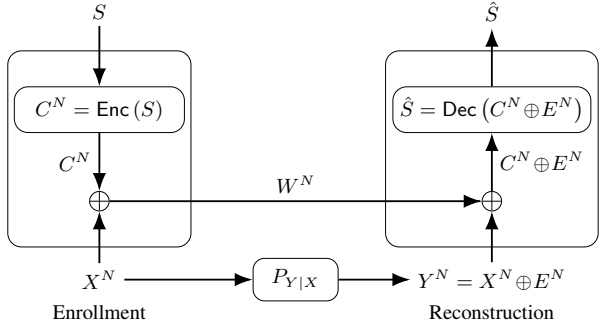

Fig.~\ref{fig:fuzzycommitment} shows the FCS, where an encoder $\Enc(\cdot)$ adds a codeword $\displaystyle C^N$, uniformly distributed over a set with cardinality $|\mathcal{S}|$, modulo-2 to the binary noiseless PUF-output sequence $\displaystyle X^N$ during enrollment. We show in Section~\ref{sec:commonsteps} that the sequence $X^N$ and its noisy version $Y^N$ can be obtained by applying the post-processing steps in Fig.~\ref{fig:postprocessing} to RO outputs $\widetilde{X}^L$ and its noisy version $\widetilde{Y}^L$, respectively. The sum $\displaystyle W^N=C^N\xor X^N$ is publicly sent through a noiseless and authenticated channel, and it is called \textit{helper data}. The modulo-2 sum of $W^N$ and the noisy PUF-output sequence $Y^N =X^N \xor E^N$, where $E^N$ is the binary error vector, gives the noisy codeword $\displaystyle C^N\xor E^N$. Using the noisy codeword, a channel decoder $\displaystyle \Dec(\cdot)$ estimates the secret key $S$ during reconstruction. A reliable secret-key agreement is possible by using $X^N$, $Y^N$, and $W^N$ \cite{AhlswedeCsiz,Maurer}.

One can achieve a (secret-key, privacy-leakage) rate pair $ (R_\text{s}\text{,}R_\ell)$ using the FCS with perfect secrecy if, given any $\epsilon\!>\!0$, there is some $N\!\geq\!1$, and an encoder and a decoder for which $\displaystyle R_\text{s}=\frac{\log|\mathcal{S}|}{N}$ and
\begin{alignat}{2}
&\Pr[S\ne\hat{S}] \leq \epsilon && (\text{reliability}) \label{eq:reliabilityconst}\\
&I\big(S;W^N\big)\!=\!0 && (\text{perfect secrecy})\label{eq:secrecyconst}\\
&\frac{1}{N}I\big(X^N;W^N\big) \leq R_\ell+\epsilon. \quad\quad\quad&&(\text{privacy})  \label{eq:privacyconst}
\end{alignat}
Condition (\ref{eq:secrecyconst}) ensures that the public side information $W^N$ does not leak any information about the secret key, so one achieves perfect secrecy. The normalized information that $W^N$ leaks about the PUF output sequence $X^N$ is considered in (\ref{eq:privacyconst}). If one should asymptotically limit the unnormalized privacy leakage $I(X^N;W^N)$, private keys available during enrollment and reconstruction are necessary \cite{IgnaTrans}, which is not realistic or practical; see the discussions in \cite{bizimWZ}.
 
Suppose the measurement channel $P_{Y|X}$ is a binary symmetric channel (BSC) with crossover probability $p$, and $X$ is independent and identically distributed (i.i.d.) according to a uniform distribution. Define $\displaystyle H_b(p)\!=\!-p\log p-(1\!-p)\log(1\!-p)$ as the binary entropy function. The region $\displaystyle \mathcal{R}$ of all achievable (secret-key, privacy-leakage) rate pairs for the FCS with perfect secrecy is \cite{IgnatenkoFuzzy}
\begin{align}
\mathcal{R}\! =\! \big\{ \left(R_\text{s},R_\ell\right)\!\colon\!\quad 0\leq R_\text{s}\leq 1-H_b(p),\quad R_\ell\geq 1\!-\!R_\text{s} \big\}.\label{eq:ls0}
\end{align}
We plot this region in Section~\ref{sec:comparisons} to evaluate the secret-key and privacy-leakage rates achieved by the proposed ECC.

The FCS is a particular realization of the CS model. The region $\mathcal{R}_{\text{cs}}$ of all achievable (secret-key, privacy-leakage) rate pairs for the CS model, where a generic encoder is used to confidentially transmit an embedded secret key to a decoder that observes $Y^N$ and the helper data $W^N$, is given in \cite{IgnaTrans, LaiTrans} as
the union over all $P_{U|X}$ of the set of achievable rate pairs $\left(R_\text{s},R_\ell\right)$ such that
\begin{align}
\Big\{0\leq R_\text{s}\leq I(U;Y),\qquad R_\ell\geq I(U;X)-I(U;Y)\!\Big\}\label{eq:chosensecret}
\end{align}
where $P_X$ is the probability distribution of $X$ and the alphabet $\mathcal{U}$ of the auxiliary random variable $U$ can be limited to have the size $\displaystyle |\mathcal{U}|\!\leq\!|\mathcal{X}|+1$ as $U-X-Y$ forms a Markov chain. The FCS achieves a boundary point of $\mathcal{R}_{\text{cs}}$ for a BSC $P_{Y|X}$ only at the point $\displaystyle (R_\text{s}^*,R_\ell^*)\!=\!(1\!-\!H_b(p),H_b(p))$. To achieve the other points on the rate-region boundary, one should use a nested code construction as in \cite{bizimWZ} or a binning based construction as in \cite{MatthieuPolar}, both of which require careful polar code \cite{Arikan} designs. This is not necessary to illustrate the gains from the new set of transforms and it suffices to combine the new set with the FCS.
\section{Post-processing Steps}\label{sec:commonsteps}
We consider a 2D array of $r\!\times\!c$ ROs. Denote the continuous-valued outputs of $L\!=\!r\!\times\!c$ ROs as the vector random variable $\widetilde{X}^L$, distributed according to $\displaystyle f_{\widetilde{X}^L}$. Suppose that the noise component $\widetilde{E}_j$ on the $j$-th RO output  is Gaussian distributed with zero mean for all $j=1,2,\ldots,L$ and that the noise components are mutually independent. Denote the noisy RO outputs as $\widetilde{Y}^L\!=\!\widetilde{X}^L\!+\!\widetilde{E}^L$. We extract binary vectors $X^N$ and $Y^N$ from $\widetilde{X}^L$ and $\widetilde{Y}^L$, respectively, and define binary error variables $\displaystyle E_i\!=\!X_i\xor Y_i$ for $i\!=\!1,2,\ldots,N$. 

\begin{figure}[t]
	\centering
	\includegraphics[width=0.485005\textwidth, height=0.5005\textheight, keepaspectratio]{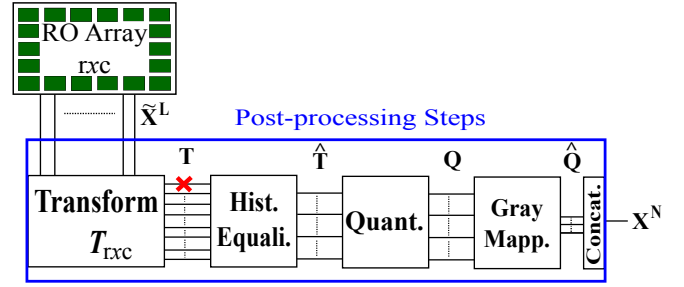}
	\caption{The transform-coding steps.} 
	\label{fig:postprocessing}
\end{figure} 

The post-processing steps used during the enrollment (and reconstruction) to extract a bit sequence $X^N$ (and its noisy version $Y^N$) are depicted in Fig.~\ref{fig:postprocessing}. These steps are transformation, histogram equalization, quantization, Gray mapping, and concatenation. Since RO outputs $\widetilde{X}^L$ are correlated, we apply a transform $\emph{T}_{r\!\times\!c}(\cdot)$ for decorrelation. We model all transform coefficients and noise components as random variables with Gaussian marginal distributions. A transform-coefficient output $T$ that comes from a distribution with mean $\mu\neq 0$ and variance $\sigma^2\neq 1$ is converted into a standard Gaussian random variable during histogram equalization, which reduces the hardware area when multiple bits are extracted. Independent bits can be extracted from transform coefficients by setting the quantization boundaries of a $K$-bit quantizer to
\begin{equation}\label{eq:quantsteps}
b_k=Q^{-1}\left(1-\dfrac{k}{2^K}\right) \text{ for } k=0,1,\dots,2^K
\end{equation}
where $Q(\cdot)$ is the $Q$-function. Quantizing a coefficient $\hat{T}$ to $k$ if $\displaystyle b_{k-1}\!<\!\hat{T}\!\leq\!b_k$ ensures that $X^N$ is uniformly distributed, which is necessary to achieve the rate point where the FCS is optimal.

One can use scalar quantizers without a performance loss in security if the RO output statistics satisfy certain constraints \cite{benimdissertation}. We do not use the first transform coefficient, i.e., DC coefficient, for bit extraction since it corresponds to the average over the RO array, known by an attacker \cite{benimdissertation}. Furthermore, Gray mapping ensures that the neighboring quantization intervals result in only one bit flip. This is a good choice as the noise components $E_i$ for all $i=1,2,\ldots,N$ have zero mean. The sequences extracted from transform coefficients are concatenated to obtain the sequence $X^N$ (or $Y^N$).

\section{New Orthogonal Transforms}\label{sec:neworth}
A useful metric to measure the complexity of a transform is the number of operations required for computations. Consider only RO arrays of sizes $ r\!=\!c\!=\!8$ and $16$, which are powers of 2, so fast algorithms are available. In \cite{benimdissertation}, the DWHT is suggested as the best candidate among the set of transforms \{DCT, DHT, KLT, DWHT\} for RO PUF applications with a low-complexity constraint such as internet of things (IoT) applications.

In \cite{bizimMDPI}, we extend an input-selection algorithm to compute the 2D $16\times16$ DWHT by applying a $2\times2$ matrix operation recursively to illustrate that the DWHT requires a small hardware area in a field programmable gate array (FPGA) since it does not require any multiplications. Following this observation, we propose a set of transforms that are orthogonal (to decorrelate the RO outputs better), that have matrix elements $1$ or $-1$ (to eliminate multiplications), and that have size of $16\times16$ (to apply the input-selection algorithm given in \cite{bizimMDPI} to further reduce complexity). We show in the next section that these transforms provide higher reliability than other transforms previously considered in the literature.

\subsection{Orthogonal Transform Construction and Selection}\label{subsec:orthtransselection}
Consider an orthogonal matrix $A$ with elements $1$ or $-1$ and of size $k\times k$, i.e., $AA^{T}= I$, where $T$ is the matrix transpose and $I$ is the identity matrix of size $k\times k$. It is straightforward to show that the following matrices are also orthogonal:
\begin{align} 
&\Biggl[ \begin{matrix}
A&A\\ A&\!-\!A
\end{matrix} \Biggr],
\Biggl[ \begin{matrix}
A&A\\ \!-\!A&A
\end{matrix} \Biggr],
\Biggl[ \begin{matrix}
A&\!-\!A\\ A&A
\end{matrix} \Biggr],
\Biggl[ \begin{matrix}
\!-\!A&A\\ A&A
\end{matrix} \Biggr],\nonumber\\
\Biggl[ &\begin{matrix}
\!-\!A&\!-\!A\\\! -\!A&A
\end{matrix} \Biggr],
\Biggl[ \begin{matrix}
\!-\!A&\!-\!A\\ A&\!-\!A
\end{matrix} \Biggr],
\Biggl[ \begin{matrix}
\!-\!A&A\\ \!-\!A&\!-\!A
\end{matrix} \Biggr],
\Biggl[ \begin{matrix}
A&\!-\!A\\ \!-\!A&\!-\!A
\end{matrix} \Biggr].\label{eq1}
\end{align} 
Since $2^{k^2}$ possible matrices should be checked for orthogonality, we choose $k\!=\!4$ to keep the complexity of the exhaustive search for orthogonal matrices low. The result of the exhaustive search is a set of orthogonal matrices $A$ of size $4\!\times\! 4$. By applying the matrix construction methods in (\ref{eq1}) twice consecutively, we obtain $12288$ unique orthogonal transforms of size $16\!\times\! 16$ with elements $1$ or $\displaystyle -1$. 

We apply these orthogonal transforms, one of which is the DWHT, to an RO dataset to select the orthogonal transform whose maximum bit-error probability over the transform coefficients is minimum. This selection method provides reliability guarantees to every transform coefficient. An ECC that has a higher code dimension than it is achievable according to the Gilbert-Varshamov (GV) bound \cite{GilbertGV,varshamovGV} for the maximum error probability over the transform coefficients of the ST, is given in Section~\ref{subsec:codeselection}. This illustrates that our selection method is conservative and the block-error probability is substantially smaller than $10^{-9}$. 

There are also other orthogonal transforms of size $16\times 16$ but we illustrate in the next section that the new set suffices to significantly increase the reliability of the extracted bits as compared to previously considered transforms and previous RO PUF methods.

\section{Performance Evaluations}\label{sec:comparisons}
We use RO arrays of size $16\!\times \!16$ from the RO dataset in \cite{ROPUF} and apply the transform-coding steps in Fig.~\ref{fig:postprocessing} to compare the previously considered transforms with the new set of transforms in terms of their reliability, uniqueness, and security. We illustrate that a Bose-Chaudhuri-Hocquenghem (BCH) code can be used for error correction in combination with the FCS to achieve a block-error probability smaller than the common value of $10^{-9}$.

\begin{figure}[t]
	\centering
%
%
\begin{tikzpicture}

\begin{axis}[%
width=7.221cm,
height=3.1cm,
at={(0cm,0cm)},
scale only axis,
xmin=0,
xmax=0.02,
xlabel style={font=\color{white!15!black}},
xlabel={Transform Coefficient Bit Error Probabilities},
scaled x ticks=false,
xticklabel style=/pgf/number format/fixed, /pgf/number format/precision=3,
ymin=0,
ymax=0.3,
axis background/.style={fill=white},
xmajorgrids,
ymajorgrids,
legend style={at={(1.046,0.99)}, anchor = north east,legend cell align=left, align=left, draw=white!15!black}
]
\addplot[ybar interval, fill=blue, fill opacity=0.990, draw=black, area legend] table[row sep=crcr] {%
x	y\\
0	0.0627450980392157\\
0.002	0.00784313725490196\\
0.004	0.0666666666666667\\
0.006	0.203921568627451\\
0.008	0.290196078431373\\
0.01	0.231372549019608\\
0.012	0.101960784313725\\
0.014	0.0352941176470588\\
0.016	0.0352941176470588\\
};
\addlegendentry{ST}

\addplot[ybar interval, fill=red, fill opacity=0.75, draw=black, area legend] table[row sep=crcr] {%
x	y\\
0	0.00784313725490196\\
0.002	0.00784313725490196\\
0.004	0.0666666666666667\\
0.006	0.247058823529412\\
0.008	0.286274509803922\\
0.01	0.215686274509804\\
0.012	0.109803921568627\\
0.014	0.0313725490196078\\
0.016	0.0156862745098039\\
0.018	0.0117647058823529\\
0.02	0.0117647058823529\\
};
\addlegendentry{DWHT}

\addplot[ybar interval, fill=violet, fill opacity=0.9, draw=black, area legend] table[row sep=crcr] {%
x	y\\
0.001	0.00392156862745098\\
0.002	0\\
0.003	0.00392156862745098\\
0.004	0.0117647058823529\\
0.005	0.0156862745098039\\
0.006	0.0470588235294118\\
0.007	0.125490196078431\\
0.008	0.223529411764706\\
0.009	0.188235294117647\\
0.01	0.156862745098039\\
0.011	0.129411764705882\\
0.012	0.0392156862745098\\
0.013	0.0352941176470588\\
0.014	0.0156862745098039\\
0.015	0.00392156862745098\\
0.016	0.00392156862745098\\
};
\addlegendentry{DCT}

\addplot [color=blue, line width=3.0pt, solid]
table[row sep=crcr]{%
	0.00886923921568628	0\\
	0.00886923921568628	0.3\\
};
\addlegendentry{ST Mean}

\addplot [color=red, line width=4.2pt, solid]
table[row sep=crcr]{%
	0.00946365490196079	0\\
	0.00946365490196079	0.3\\
};
\addlegendentry{DWHT Mean}

\addplot [color=violet, line width=3.6pt, solid]
  table[row sep=crcr]{%
0.0094953137254902	0\\
0.0094953137254902	0.3\\
};
\addlegendentry{DCT Mean}

\end{axis}
\end{tikzpicture}%
	\caption{The histograms and means of the bit-error probabilities of the transform coeeficients obtained from the DCT, DWHT, and the selected transform (ST) from the new set.} 
	\label{fig:BERComparisonsofTrans}
\end{figure}
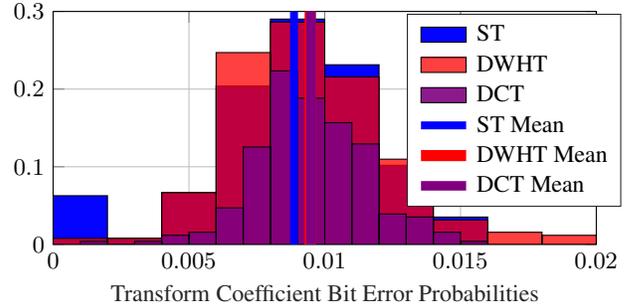

\subsection{Transform Comparisons}
We compare the orthogonal transform selected from the new set, i.e., the ST, with the DCT and DWHT in terms of the bit-error probabilities of the $255$ transform coefficients obtained from the RO dataset in \cite{ROPUF}. Fig.~\ref{fig:BERComparisonsofTrans} illustrates the bit-error probabilities of the DCT, DWHT, and the ST. The mean of the ST is smaller than the means of the DCT and DWHT. Furthermore, the maximum bit-error probability of the DCT and ST are almost equal and are less than the maximum error probability of the DWHT. Most importantly, the ST has a large set of transform coefficients with bit-error probabilities close to zero, so an ECC design for the maximum or mean bit-error probability of the ST would give pessimistic rate results. We propose in the next section an ECC for  the ST to achieve a smaller block-error probability than the block-error probability for the DCT.

\subsection{Uniqueness and Security}\label{subsec:uniqueness}

A common measure to check the randomness of a bit sequence is uniqueness, i.e., the average fractional Hamming distance (HD) between the sequences extracted from different RO PUFs \cite{bizimpaper}. The rate region in (\ref{eq:ls0}) is valid if the extracted bit sequences are uniformly distributed, making the uniqueness a valid measure for the FCS.

Uniqueness results for the DCT, DWHT, KLT, and DHT have a mean HD of $0.5000$ and HD variances of approximately $\displaystyle 7\!\times \!10^{-4}$ \cite{bizimMDPI}, which are close to optimal and better than previous RO PUF results. For the ST, we obtain a mean HD of $0.5001$ and a HD variance of $\displaystyle 2.69\!\times \!10^{-2}$. This suggests that the ST has good average uniqueness performance, but there might be a small set of RO PUFs from which slightly biased bit sequences are extracted. The latter can be avoided during manufacturing by considering uniqueness as a parameter in yield analysis of the chip that embodies the PUF. We apply the national institute of standards and technology (NIST) randomness tests \cite{NIST} to check whether there is a detectable deviation from the uniform distribution in the sequences extracted by using the ST. The bit sequences generated with the ST pass most of the randomness tests, which is considered to be an acceptable result \cite{NIST}. A correlation thresholding approach in \cite{bizimtemperature} further improves security.

\subsection{Code Selection}\label{subsec:codeselection}
Consider the scenario where secret keys are used as an input to the advanced encryption standard (AES), a  symmetric-key cryptosystem, with a key size of $128$ bits, so the code dimension of the ECC should be at least $128$ bits. The maximum error probability over the transform coefficients of the ST is $p_{\text{max}}=0.0149$, as shown in Fig.~\ref{fig:BERComparisonsofTrans}. Furthermore, assume that we use an ECC with a bounded minimum distance decoder (BMDD) to keep the complexity low. A BMDD can correct all error patterns with up to $\lfloor\frac{d_{\text{min}-1}}{2}\rfloor$ errors, where $d_{\text{min}}$ is the minimum distance of the code. It is straightforward to show that the ECC should have at least a minimum distance of $d_{\text{min}}=41$ to achieve a block-error probability of $P_\text{B}\leq 10^{-9}$ if all transform coefficients are assumed to have a bit-error probability of $p_{\text{max}}$. None of binary BCH and Reed-Solomon (RS) codes, which have good minimum-distance properties, can satisfy these parameters. Similarly, the GV bound computed for $p_{\text{max}}$ shows that there exists a linear binary ECC with code dimension $98$. Consider the  binary BCH code with the block length $255$, code dimension $131$ that is greater than the code dimension of $98$ given by the GV bound, and minimum distance $\displaystyle d_{\text{min,BCH}}=37$ that is close to the required value of $d_{\text{min}}=41$. We illustrate in the next section that this BCH code provides a block-error probability significantly smaller than $10^{-9}$. 

\subsection{Reliability, Privacy, and Secrecy Analysis of the Code}
We now show that the proposed ECC satisfies the block-error probability constraint. The block-error probability $P_\text{B}$ for the $\text{BCH}(255,131,37)$ code with a BMDD is equal to the probability of having more than $18$ errors in the codeword, i.e., we have
\begin{align}
 P_\text{B} = \sum_{j=19}^{255}\Bigg[\sum_{\mathcal{D}\in\mathcal{F}_j}\prod_{i\in \mathcal{D}}p_{i}\,\bigcdot\prod_{i\in \mathcal{D}^{c}}(1-p_{i}) \Bigg] \label{eq:blockerrorforbch}
\end{align}
where $p_{i}\leq p_{\text{max}}$ is the bit-error probability of the $i$-th transform coefficient, as in Fig.~\ref{fig:BERComparisonsofTrans}, for $i\!=\!2,3,\ldots,256$, $\displaystyle \mathcal{F}_j$ is the set of all size-$j$ subsets of the set $\displaystyle\{2,3,\ldots,256\}$, and $\mathcal{D}^{c}$ denotes the complement of the set $\mathcal{D}$. The  bit-error probabilities $p_{i}$ represent probabilities of independent events due to the mutual independence assumption for transform coefficients and one-bit quantizers used. 

The evaluation of (\ref{eq:blockerrorforbch}) requires $ \sum_{j=0}^{18}{255\choose j}\approx 1.90\!\times\!10^{27}$ different calculations, which is not practical. We therefore apply the discrete Fourier transform - characteristic function (DFT-CF) method \cite{DFTCF} to (\ref{eq:blockerrorforbch}) and obtain the result $P_\text{B}\!\approx\!2.860\!\times\!10^{-12}\!<\!10^{-9}$. This value is smaller than the block-error probabilitiy $P_{\text{B,DCT}}= 1.26\times 10^{-11}$ obtained in \cite{benimdissertation} for the DCT with the same code. The block-error probability constraint is thus satisfied by using the $\text{BCH}$ code although the conservative analysis suggests otherwise.

\begin{figure}[t]
	\centering
	\input{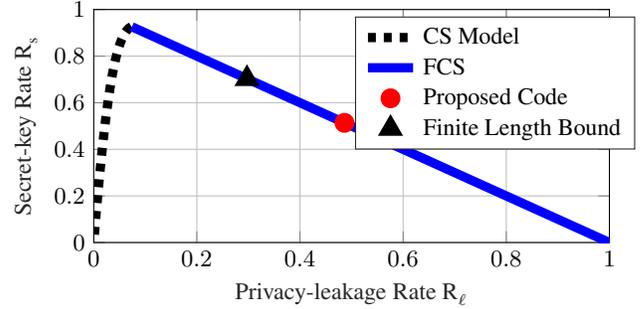}
	\caption{Boundaries of asymptotically achievable rate regions for the CS model and the FCS, operation point of the proposed code, and a finite-length bound for $N=255$ bits and $P_\text{B}=10^{-9}$.} 
	\label{fig:ratecomparison}
\end{figure}

The rate regions given in (\ref{eq:ls0}) and (\ref{eq:chosensecret}) are asymptotic results, i.e., they assume $N\rightarrow \infty$. Since separate channel and secrecy coding is optimal for the FCS, we can use the finite length bounds for a BSC $P_{Y|X}$ with crossover probability $p\!=\! \frac{1}{L-1}\sum_{i=2}^Lp_{i}\!\approx\!0.0088$, i.e., the error probability averaged over all used coefficients. In \cite{benimdissertation}, we show that the $\text{BCH}(255,131,37)$ code achieves $(R_{\text{s,BCH}},R_{\ell,\text{BCH}})\approx(0.514,\,0.486)$ bits/source-bit, significantly better than previously proposed codes in the RO PUF literature, so it suffices to compare the proposed code with the best possible finite-length results for the FCS. We use Mrs. Gerber's lemma \cite{WZ}, giving the optimal auxiliary random variable $U$ in (\ref{eq:chosensecret}), to compute all points in the region $\mathcal{R}_{\text{cs}}$. We plot all achievable rate pairs, the (secret-key, privacy-leakage) rate pair of the proposed BCH code, and a finite-length bound for the block length of $N=255$ bits and $P_\text{B}\!=\!10^{-9}$ in Fig.~\ref{fig:ratecomparison}.

The maximum secret-key rate is $R_\text{s}^*\!\approx\!0.9268$ bits/source-bit with a corresponding minimum privacy-leakage rate of $R_\ell^*\!\approx\!0.0732$ bits/source-bit. The gap between the points $(R_{\text{s,BCH}},R_{\ell,\text{BCH}})$ and $(R_{\text{s}}^*,R_\ell^*)$ can be partially explained by the short block length of the code and the small block-error probability. The finite-length bound given in \cite[Theorem 52]{Polyanskiy} shows that the rate pair $(R_\text{s},R_\ell)\!=\!(0.7029,0.2971)$ bits/source-bit is achievable by using the FCS, as depicted in Fig.~\ref{fig:ratecomparison}. One can thus improve the rate pairs by using better codes and decoders with higher hardware complexity, which is undesirable for IoT applications. Fig.~\ref{fig:ratecomparison} also illustrates the fact that there are operation points of the region $\mathcal{R}_{\text{cs}}$ that cannot be achieved by using the FCS and, e.g., a nested polar code construction from \cite{bizimWZ} should be used to achieve all points in $\mathcal{R}_{\text{cs}}$.

\section{Conclusion}\label{sec:conclusion}
We proposed a new set of transforms that are orthogonal (so that the decorrelation efficiency is high), that have elements $1$ or $-1$ (so that the hardware complexity is low), and that have a size of $k\times k$ where $k$ is a power of 2 (so that an input-selection algorithm can be applied to further decrease complexity). By using one-bit uniform quantizers for each transform coefficient obtained by applying the ST, we obtained bit-error probabilities that are on average smaller than the bit-error probabilities obtained from previously considered transforms. We proposed a BCH code as the ECC for RO PUFs in combination with the FCS. This code achieves the best rate pair in the RO PUF literature and it gives a block-error probability for the ST that is substantially smaller than for the DCT. We illustrated that the FCS cannot achieve all possible rate points. In future work, in combination with the new set of transforms, we will apply a joint vector quantization and error correction method by using nested polar codes to achieve rate pairs that cannot be achieved by the FCS.


\bibliographystyle{IEEEbib}
\bibliography{references}

\end{document}